\documentclass[journal]{IEEEtran}
\IEEEoverridecommandlockouts
\usepackage{cite}
\usepackage{amsmath,amssymb,amsfonts}
\usepackage{algorithmic}
\usepackage{graphicx}
\usepackage{textcomp}
\usepackage{xcolor}
\usepackage[british,UKenglish,USenglish,english,american]{babel}
\usepackage[utf8]{inputenc}
\usepackage{hyperref}
\usepackage{tabularx}
\usepackage{colortbl}
\usepackage{multirow}
\usepackage{xcolor}
\usepackage[normalem]{ulem}
\usepackage{listings}
\useunder{\uline}{\ul}{}

\usepackage[switch]{lineno}
\usepackage{longtable}
\usepackage{pdflscape}
\usepackage{framed}
\usepackage{float}

\newcommand{\etal}[0] {\textit{et al.}}

\usepackage[normalem]{ulem} 

\def\BibTeX{{\rm B\kern-.05em{\sc i\kern-.025em b}\kern-.08em
    T\kern-.1667em\lower.7ex\hbox{E}\kern-.125emX}}
    
\begin{document}


\title{Framework to coordinate ubiquitous devices with SOA standards}

\author{
\IEEEauthorblockN{Oscar A. Testa\IEEEauthorrefmark{1}, 
Efraín R. Fonseca C.\IEEEauthorrefmark{2}, 
Germán Montejano\IEEEauthorrefmark{3}, 
and Oscar Dieste\IEEEauthorrefmark{4}} \\
\IEEEauthorblockA{\IEEEauthorrefmark{1}\textit{Facultad de Ciencias Exactas y Naturales,} 
\textit{Universidad Nacional de La Pampa.}
Santa Rosa, Argentina. Email: otesta@exactas.unlpam.edu.ar} \\
\IEEEauthorblockA{\IEEEauthorrefmark{2}\textit{Departamento de Ciencias de la Computación,} 
\textit{Universidad de las Fuerzas Armadas ESPE.}
Quito, Ecuador. Email: erfonseca@espe.edu.ec} \\
\IEEEauthorblockA{\IEEEauthorrefmark{3}\textit{Facultad de Ciencias Físico Matemáticas y Naturales,} 
\textit{Universidad Nacional de San Luis.}
San Luis, Argentina. Email: german.a.montejano@gmail.com} \\
\IEEEauthorblockA{\IEEEauthorrefmark{4}\textit{Escuela Técnica Superior de Ingenieros Informáticos,} 
\textit{Universidad Politécnica de Madrid.}
Madrid, España. Email: odieste@fi.upm.es}
}


\maketitle

\IEEEpubidadjcol

\begin{abstract}

\textbf{\textit{Context:}}  
Ubiquitous devices and pervasive environments are in permanent interaction in people's daily lives. In today's hyper-connected environments, it is necessary for these devices to interact with each other, transparently to the users. The problem is analyzed from the different perspectives that compose it: SOA, service composition, interaction, and the capabilities of ubiquitous devices.

\textbf{\textit{Problem:}}  
Currently, ubiquitous devices can interact in a limited way due to the proprietary mechanisms and protocols available on the market. The few proposals from academia have hardly achieved an impact in practice. This is not in harmony with the situation of the Internet environment and web services, which have standardized mechanisms for service composition.

\textbf{\textit{Aim:}}  
Apply the principles of SOA, currently standardized and tested in the information systems industry, for the connectivity of ubiquitous devices in pervasive environments. For this, a coordination framework based on these technologies is proposed.

\textbf{\textit{Methodology:}}  
We apply an adaptation of Design Science in our environment to allow the iterative construction and evaluation of prototypes. For this, a proof of concept is developed on which this methodology and its cycles are based.

\textbf{\textit{Results:}}  
We built and put into operation a coordination framework for ubiquitous devices based on WS-CDL, along with a proof of concept. In addition, we contribute to the WS-CDL language in order to support the characteristics of specific ubiquitous devices.

\end{abstract}



\section{Introduction}\label{sec:introduction}

Mark Weiser introduced the term ubiquitous computing (he also used the acronym \textit{ubicomp})  \cite{weiser_ubiquituouscomputing} in the late '80s and early '90s. 
Ubiquitous computing is a technology that claims computers are not seen as such so that they can be integrated into people's lives in a transparent way \cite{weiser_ubiquituouscomputing} \cite{VASSEUR20103}. Ubiquitous devices are electronic devices with processing and communication capabilities and can be found anywhere: in the office, in the car, in the house, or even in the clothes we wear.

Technological advances have allowed ubiquitous devices to be both generators and consumers of services \cite{VASSEUR20103};this results in an environment of inter-device cooperation, which allows for the composition of more complex functionalities. By composition, we understand how ubiquitous devices can be combined to perform a given task \cite{compositionservicesdivideandconquer}. The composition implies that devices must communicate with each other to obtain a value-added service, which in turn brings challenges with it, such as heterogeneity, connectivity, fault tolerance, security, and customization of the devices, e.g., services provision according to user's preferences. Since ubiquitous devices have resource limitations (e.g., limited memory and battery), special considerations should be made regarding the performance and efficiency of device composition. All these difficulties make the composition of ubiquitous devices an important area of research, where progress is not clear to this day \cite{Webservicescomposition_decade_overview}.

The communication between ubiquitous devices is not usually carried out using standardized mechanisms; 
instead, most protocols are proprietary, e.g., alarm systems, communications between vehicles of a specific brand, 
household appliances, etc. This represents a significant limitation in the composition 
of ubiquitous devices \cite{VASSEUR20103} \cite{Webservicescomposition_decade_overview}, 
as it prevents devices from different manufacturers from interacting with each other or, alternatively, 
requires devices to implement multiple protocols (consider, for example, devices that simultaneously operate under Google Home%
\footnote{\url{https://home.google.com/welcome/}}%
 and Alexa%
 \footnote{\url{https://www.alexa.com}}%
), with the associated design complexities and costs this entails.

In contrast to ubiquitous devices, other areas have standardized coordination mechanisms. An evident example is the Internet, where the TCP/IP protocol ensures communication between different nodes in the network. Other technologies that are de facto standards today

The SOA (Service-Oriented Architecture) architecture represents a successful approach for creating applications through the composition of components that communicate with each other (using, in fact, the aforementioned technologies). The characteristics of SOA (heterogeneity and interoperability of services, hardware independence, flexible coupling, etc.) prompted us to bring together the existing specifications and standards in SOA for coordinating ubiquitous devices. If each ubiquitous device is considered as a provider or consumer of a service, ubiquitous devices fit perfectly into the SOA architecture. The advantages of adopting an SOA-like architecture in the ubiquitous world are numerous: standardization, transparency, and interoperability with other platforms. In contrast, the cost is quite low.

We have verified that it is possible to execute WSCDL choreographies \cite{8966390} \cite{9067311} on ubiquitous devices with very limited resources. To achieve this, we have developed and adapted various existing mechanisms in SOA to be executable on low-performance devices, such as ubiquitous devices.

Our contributions are as follows:
\begin{itemize}
\item We have developed a framework (with versions in C++ and PHP) to coordinate ubiquitous devices using WSCDL choreographies.
\item We have made adaptations to the WSCDL standard to address common situations in ubiquitous devices, such as device disappearance and transaction handling.
\item We have implemented a proof of concept using devices with memory and processing limitations, such as Arduino boards and Raspberry Pi B+.
\end{itemize}

The rest of the article is organized as follows: Section 2 describes the background and state of the art of ubiquitous systems, emphasizing coordination and SOA as a service coordination and composition mechanism. Section 3 details the selected research methodology and research aims. Section 4 describes the coordination framework developed. A concept test carried out to implement the developed framework is described in Section 5. Finally, Section 6 presents the conclusions and discussion. Section 7 presents the annexes, where the most important classes of both the framework and the proof of concept are shown.

\section{Background}\label{sec:background}

\subsection{Ubiquitous computing}

Ubiquitous computing is a technological development aimed at making computers imperceptible in the environment as distinct objects, enabling their use by humans to be as transparent and convenient as possible, thereby facilitating integration into daily life. Ubiquitous devices are all those devices that can exist everywhere; in other words, they are electronic devices with processing and communication capabilities that can be found anywhere: the office, the car, or even the clothes we wear \cite{weiser_ubiquituouscomputing}.

For several years, ubiquitous devices have gained importance in people's daily lives, mainly because they feature different types of sensors (positioning, proximity, luminosity, temperature, etc.), facilitate connectivity even in areas with weak signal or limited network access, enable technological convergence (computing, media, telephony, etc.), and provide access to various services (maps, assistance, etc.) \cite{zhao2021handbook}. An example of everyday use is the ubiquitous devices available in a car, which

\subsection{Pervasive systems}

The objective of ubiquitous computing is to allow users to access services and resources anytime, irrespective of their location. Pervasive computing is a related concept aiming to provide spontaneous services created on the fly \cite{10.1145/1387249.1387256}.

Pervasive environments are characterized by being populated with numerous ubiquitous devices that, thanks to the extreme integration of electronic elements, are invisible to the user and constantly track human activity \cite{zhao2021handbook} \cite{weiser_ubiquituouscomputing}. An example of this is the use of location-based services, where areas of art, commerce, etc., are offered to us as we move through a given place.

Pervasive systems are composed of ubiquituous devices with limited storage and processing power. Devices are autonomous, heterogeneous, and typically embedded in larger systems; consequently, they become invisible to their users. Devices use wireless connections and open protocols to communicate with other devices or systems and achieve their objectives.

Hardware evolution makes pervasive devices improve their processing power (e.g., RaspberryPi  Zero 2 W with 1GHz quad-core 64-bit Arm Cortex-A53 CPU and 512MB of SDRAM), but certain limitations are not easy to solve and still exist, e.g., the battery's useful life, availability of connection to networks, etc. However, the biggest challenge is that the existing communication protocols do not align with pervasive systems and need improvement, especially to allow device mobility \cite{ubiquitousPoslandStefan2009} \cite{weiser_ubiquituouscomputing}. 

Pervasive systems have particular characteristics compared to similar systems, such as the Internet of Things (IoT). The most distinctive feature is that IoT focuses more on connecting the devices, and pervasive computing focuses more on human-computer interaction issues.

\subsection{Service composition}

Composition refers to the way in which an indefinite number of elements can be combined to perform a task \cite{Webservicescomposition_decade_overview}. The composition of different elements (services, devices, etc.) into a higher-level abstract entity aims primarily to enable that entity to perform a complex task through the use of various simpler elements \cite{Webservicescomposition_decade_overview}.

In order to carry out the composition of services or devices, it is necessary to perform certain activities (definition, selection, implementation, execution), which must follow a more or less strict sequence. Each of these phases requires a series of requirements to be met. In the \textit{definition stage}, the requirements that the composition must fulfill (orchestrations and choreographies) are established; during the \textit{selection stage}, the services or devices that will be part of the composition must be selected (which can be static, dynamic, manual, automatic, etc.); during the \textit{implementation stage}, the composition is implemented with the specific services to enable their interaction and invocation; and finally, the \textit{execution stage} is where the composition is instantiated and executed by a specific execution engine \cite{Webservicescomposition_decade_overview}.

\subsection{Service composition in pervasive systems}

Composition in pervasive environments implies that ubiquitous devices must communicate with each other in order to share the services they offer, with the aim of obtaining a value-added service that none of the participating devices can offer separately. There are several ways to achieve this goal:
\begin{itemize}

\item \textbf{Environments}, such as \textit{Aura}\cite{auraproject}, \textit{Gaia} \cite{gaiaproject}, or \textit{Oxigen} \cite{oxygenproject}. These are an abstraction layer that provides, in addition to a transparent integration mechanism for ubiquitous devices for users and developers, some form of reactivity to the environment, as pervasive computing advocates. e.g., opening the blinds when the user arrives home during the day, turning on the lights if it is nighttime, etc. In this example, we see the system reacting to an event, the user's entry and the time of day, with the connectivity of various sensors and activators (light sensor and ambient light activator).

\item \textbf{Frameworks}, such as \textit{Amigo}\cite{AmigoProjectOpenSource} or \textit{OSGi} \cite{10.1145/1402521.1402526}. Although designed in the area of intelligent agents, \textit{Jade} \cite{bellifemine} could be categorized within this group. The frameworks provide a layer of middleware that allows the integration of heterogeneous devices and services into a network or distributed environment. Users may or may not be aware of the existence of ubiquitous devices depending on the type of system and device, but developers are fully aware. Frameworks provide the foundation for communication between ubiquitous devices, but coordination must be performed programmatically.

\item \textbf{Protocols}, such as MQTT (Message Queue Telemetry Transport) \cite{mqtthomepage} or Constrained Application Protocol (CoAP) \cite{coaphomepage}. In this case, the pervasive system is a collection of independent devices that must explicitly coordinate using the primitives offered by the protocols. Composition based on these protocols is done under a subscription-publishing or request-response model, which allows devices to communicate with each other \cite{mqtthomepage,coaphomepage}.

\end{itemize}

\subsection{Service-oriented architecture}

The service-oriented architecture (SOA) allows the creation of scalable software systems that reflect an organization's business and also provides a well-defined way to expose and invoke services, facilitating the integration and interaction of both own and third-party systems.

Web Services have become the preferred technology for SOA implementation. The success of this technology lies in basing its development on existing infrastructures such as HTTP, SOAP, and XML \cite{Georgakopoulos-2008-Service-Oriented-Computing}. 
The World Wide Web Consortium (W3C) defines a web service as a software system designed to support interoperable machine-to-machine (M2M) interaction across a network \cite{webservicearchitecture}.  Please notice that CoAP \cite{coaphomepage}, mentioned above, also matches this definition.

A web service performs a specific task described by a service specification in a standard XML notation called WSDL (Web Services Description Language) \cite{ibmwebservicedefinition}. The WSDL provides the details necessary to interact with the service, hiding implementation details\cite{ibmwebservicedefinition}. The Simple Object Access Protocol (SOAP) provides the protocol for service messaging. At the beginning of 2000, REST (Representational State Transfer), a new way of implementing web services, appeared \cite{inproceedingsREST}. REST is supported, like SOA, in already existing infrastructure, such as XML and HTTP, but without using the SOAP layer.

\subsection{SOA protocols}

\begin{figure}[htbp!]
\centering
\includegraphics[scale=0.3]
{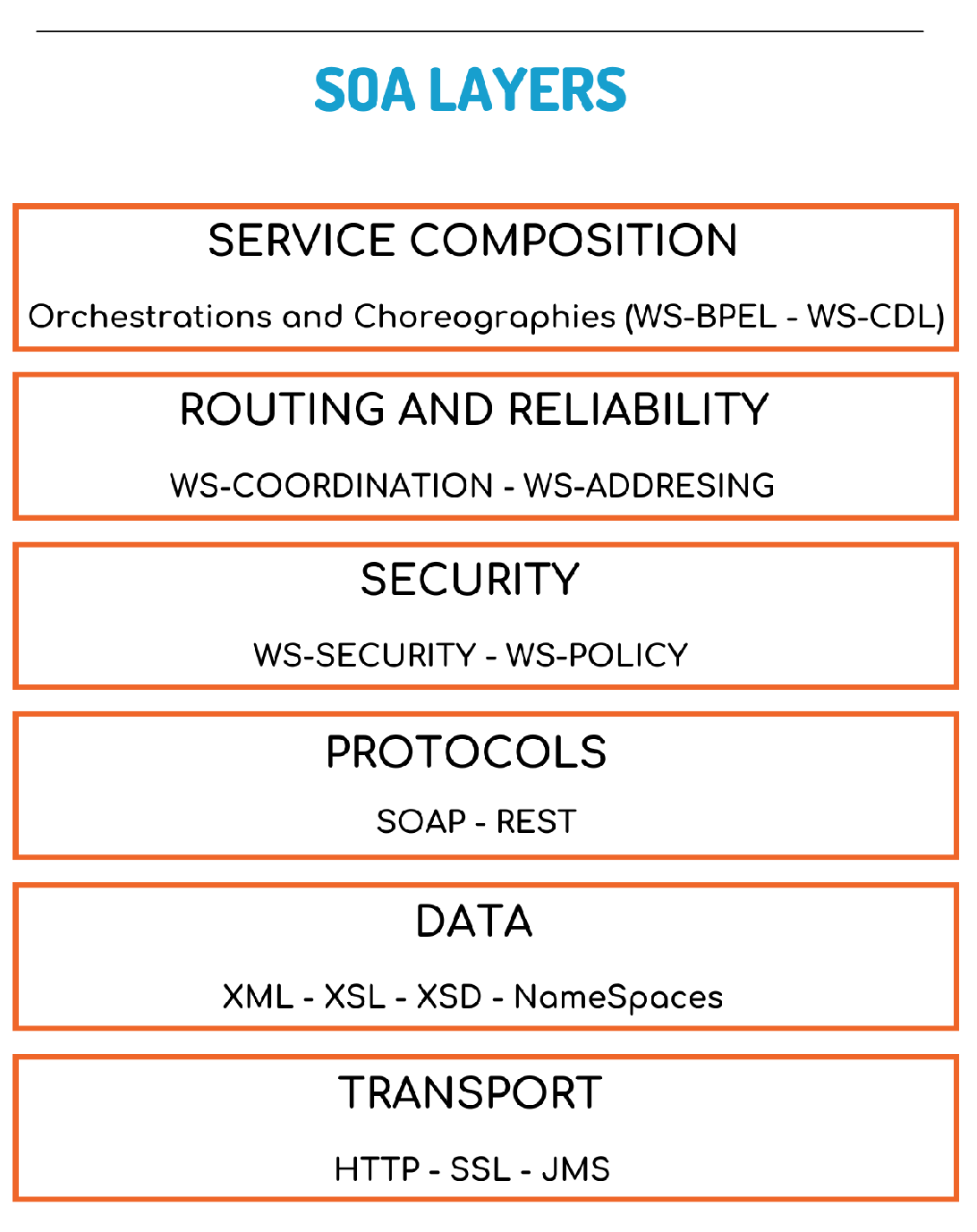}
\caption{Layers that make up SOA.}
\label{f2_5}
\end{figure}

As shown in Figure~\ref{f2_5}, there is a complex protocol stack, inspired by the OSI model of networks, that enables the open and standardized execution of complex tasks. Within this stack, there is a range of protocols that cover all aspects of SOA. First, there is the service composition protocol layer, which indicates how the different services are combined to carry out a specific task. Second, we have the coordination and reliability protocols, which, as the word implies, provide us with the capabilities of data integrity and accuracy in the exchange of information through them. Next, we have the security layer, which allows us to protect the integrity, confidentiality, and authenticity of the messages. Then, there are the protocols by which services can be invoked, such as SOAP or REST. Finally, there is the transport layer, which involves communication protocols.

\subsection{Composition in SOA}

The composition of web services is an aggregation process of several services into one, with the aim that such service can perform a more complex task, which in turn means having better applications \cite{Georgakopoulos-2008-Service-Oriented-Computing}.
There are two main composition strategies:

\begin{itemize}

\item \textbf{Orchestration}: One participant plays the role of the central coordinator. The remaining participants follow the instructions provided by the coordinator to deliver the required composite service. Orchestrations describe how services interact in a workflow, including the business logic and the logical order of interactions \cite{Rosen-2008-Applied-SOA}.

\item \textbf{Choreography}: This strategy is analogous to ballet, where the dancers perform a choreography coordinating with each other, without a director telling them what to do at every moment. Choreographies focus on the exchange of messages from the perspective of an external observer. Each service involved in the choreography must decide when to execute its operations and how to interact with other services \cite{Rosen-2008-Applied-SOA}.  

\end{itemize}

There are specifications for the implementation of choreographies and orchestrations, which have been addressed by two international organizations: OASIS (a non-profit consortium that promotes the development, convergence, and adoption of open standards for the global information society) and W3C.

The composition through orchestrations is well-defined, and there are many implementations, with BPEL or WS-BPEL being the orchestration languages standardized by OASIS \cite{Rosen-2008-Applied-SOA}.

The situation regarding choreographies is different. Currently, there is only the WS-CDL (Web Service Choreography Description Language) specification \cite{wscdl-prime:2006}, and WS-CI (Web Service Choreography Interface), an XML-based interface description language that describes the flow of messages exchanged by a Web Service participating in choreographed interactions with other services. These, however, were standardized by W3C.

\subsection{Web Service Choreography Description Language}

WS-CDL is a specification developed by the W3C Web Services Choreography Working Group \cite{wscdl-prime:2006}. The group's work concluded in July 2005, proposing WS-CDL as a "Candidate Recommendation." However, the W3C has not promoted this specification in recent years%
\footnote{The BPMN 2.0 business process specification \cite{bpmn20} has started to provide support for handling choreographies, but not based on this specification. At the time of writing, there are no industrial implementations of this specification.}%
, unlike the WS-BPEL specification for orchestrations, which is widely used in the definition of business processes.

WS-CDL defines business processes based on collaboration protocols among participants that expose their functionality through Web services. The services act as peers, and the interactions can have a lifecycle and long-lasting states. Any participant can be part of the composition, regardless of the platform or programming model used for the service implementation.

The perspective of message exchange between participants is global; that is, it does not depend on a specific point of view as in the case of orchestrations. WS-CDL provides rules that each participant uses to analyze the state of the choreography and deduce the potential next message exchange.  

Choreographies must include at least a set of roles defined by certain behaviors, a series of relationships between these roles, channels used by the roles to interact, and a space or block where the interactions defining the choreography itself are specified.  
An example of a WS-CDL choreography is shown in Listing~\ref{fig:ejemplo-coreografia}.

The example illustrates the interaction between the devices \texttt{VehiculoAccidentadoRole} and \texttt{BalizaRole}, which are specific roles declared in the corresponding section. The latter is further related to another device, \texttt{CentralBalizaRole}. This coordination is evident in the two defined interactions: \textit{``reportarAccidente''} and \textit{``publicarAccidente''}. These interactions represent the global-level messages that must be exchanged among the participants in the choreography, defining how they are related. This choreography is described and explained in detail in Section~\ref{sec:proof-of-concept}.

\begin{center}
\begin{lstlisting}[language=c++ , caption={Choreography example}, label={fig:ejemplo-coreografia}]
<interaction name="reportarAccidente" operation="informarIncidente" >
	<participate relationshipType="tns:Vehiculo_Baliza" fromRole="tns:VehiculoAccidentadoRole" toRole="tns:BalizaRole" />
	<exchange action="request" name="informarIncidente" informationType="tns:avisoIncidenteType">
		<send variable="cdl:getVariable(tns:DatosIncidente,VehiculoAccidentadoRole)"/>
		<receive variable="cdl:getVariable(tns:DatosIncidente,BalizaRole)"/>
	</exchange>
</interaction>

<interaction name="publicarAccidente" operation="publicarIncidente">
	<participate relationshipType="tns:Baliza_CentralBaliza" fromRole="tns:BalizaRole" toRole="tns:CentralBalizasRole" />
	<exchange action="request" name="informarIncidente" informationType="tns:avisoIncidenteType">
		<send variable="cdl:getVariable(tns:DatosIncidente,BalizaRole)"/>
		<receive variable="cdl:getVariable(tns:DatosIncidente,CentralBalizasRole)"/>
	</exchange>
</interaction>
\end{lstlisting}
\end{center}

\section{RESEARCH AIM \& METHODOLOGY}

While considerable progress has been made in the composition and integration of ubiquitous devices, todavía existen algunas carencias. Proposals that allow more transparent integration, e.g., \textit{Aura} \cite{auraproject} or \textit{Amigo} \cite{AmigoProjectOpenSource}, are not supported by a standard specification. Standard protocols such as MQTT \cite{mqtthomepage} or CoAP \cite{coaphomepage} delegate the specification of the composition to the programmer, with the complexity and risks this entails.  

On the other hand, all composition proposals downplay the fact that pervasive systems must interact with other systems, e.g., enterprise information systems, which are developed using other technologies, e.g., SOA. The convergence of these technologies can be achieved through the use of SOA standards, integrating ubiquitous or pervasive devices into them so that they can be seen as service providers or consumers.

Pervasive systems are challenging but share some features with other well-known software ecosystems. More concretely, if we visualize each ubiquitous device in a pervasive environment as a resource provider or consumer, the coordination of devices fits perfectly with SOA's services composition. This fact is seldom surprising: SOA is technology-independent, although it has been primarily applied to web-based systems. As long as SOA's principles apply to pervasive systems, convergence is possible.

\subsection{Research aim}

\textbf{We aim to extend the existing SOA specifications} for coordinating devices in pervasive environments. The idea has been introduced previously in the literature. Sheng \etal~\cite{Webservicescomposition_decade_overview} mentions the need for more research in this area. 

There is no work that has proposed transferring SOA concepts to the domain of ubiquitous devices. However, some authors have suggested different approaches and similar scenarios to address the issue of device coordination.  

Dragoni et al. \cite{Dragoni2017} define microservices as independent, cohesive processes that interact with others through messages. From a technical perspective, microservices must be independent components, conceptually implemented in isolation.

The same author defines microservice architecture as a new programming paradigm based on the composition of small services, each executing its own processes and communicating through lightweight mechanisms (messages). This approach is based on SOA concepts, to the extent that Netflix adopted a very similar architecture under the name Fine-grained SOA \cite{NetflixFinegrained}.

Without going into much more detail about microservices and the associated architecture, Dragoni et al. \cite{Dragoni2017} note that the way to compose microservices is primarily based on orchestrations and choreographies. Choreographies are considered more suitable, as in these, all components have the same level of relevance and importance in the execution. These concepts are based on the standards of WS-BPEL and WS-CDL, respectively, which are derived from SOA.

There are works and research conducted on service composition in other areas, such as embedded systems, actuators, or sensor networks, as in the works of \cite{6249331}, \cite{1506.02531}, \cite{5474694}, \cite{ZHOU2018299}, where they highlight the need for non-hierarchical connectivity, emphasizing the advantages of other forms of connectivity, such as choreographies.

\subsection{Research strategy}

SOA's services composition would bring multiple advantages to pervasive systems. On the one hand, SOA's architecture and programming model is standard, thus bypassing the ad-hoc, often incompatible solutions proposed in the literature. On the other, the inclusion of ubiquitous devices in the SOA protocol stack \footnote{We refer to the different layers that make up SOA technology: composition, routing and reliability, security, protocols, data and transport.}, give the possibility that ubiquitous devices stop being mere data collectors and become actors that are part of the organization.

\textbf{However, SOA's concepts cannot be directly applied}.  Composition in pervasive environments must address various device contingencies, such as memory availability, battery duration, and network access, i.e., the availability of devices cannot be guaranteed, so orchestrations and choreographies would likely fail in realistic scenarios \cite{Webservicescomposition_decade_overview}.

Achieving convergence between SOA and pervasive systems \textbf{requires harmonizing both technologies' characteristics}. For instance,  pervasive devices' memory limitations prevent using \textit{standard} SOA resources, e.g., the \textit{Apache web server} for implementing the \textit{http} protocol. However, \textit{Apache} is not the only existing \textit{http} server implementation. There are different implementations targeted to different platforms and usage contexts, e.g., \textit{Lighttpd}%
\footnote{https://www.lighttpd.net}%
. Alternatively, specific web servers could be developed for devices that do not have an operating system to install already-tested solutions or do not have enough memory to run complex programs. Another alternative to this approach is the use of restricted REST, or also the use of other protocols, such as the already mentioned MQTT. These lighter protocols, based on HTTP, are more suitable for more constrained networks, where packet size is limited or where packet loss is more likely.

\subsection{Research methodology}

We have applied the \textit{Design Science} methodology to conduct the research. Design Science creates and evaluates information technology artifacts to solve properly identified problems \cite{March:1995:DNS:222827.222832}. 
The methodology's iterative approach allows the convergence between SOA and pervasive systems to occur gradually by building sequences of prototypes that incrementally approach SOA and pervasive device worlds. La secuencia de prototipos está determinada por la list of characteristics mostrada in Table \ref{t1_5}. We have used a typical working domain of pervasive systems, described in Section~\ref{sec:proof-of-concept}, so that the prototypes correspond to realistic scenarios. Finally, we provide a proof of concept for our proposal.

\section{Proposed coordination framework}
\label{sec:framework}

First, we will describe the general architecture of the proposed coordination framework. A continuación, indicaremos how the limitations of ubiquitous devices could be technically managed to converge with SOA concepts and technologies.

\subsection{Framework architecture}

\subsubsection{Transition loop}

\begin{figure}[]
\centering
\includegraphics[width=\linewidth]
{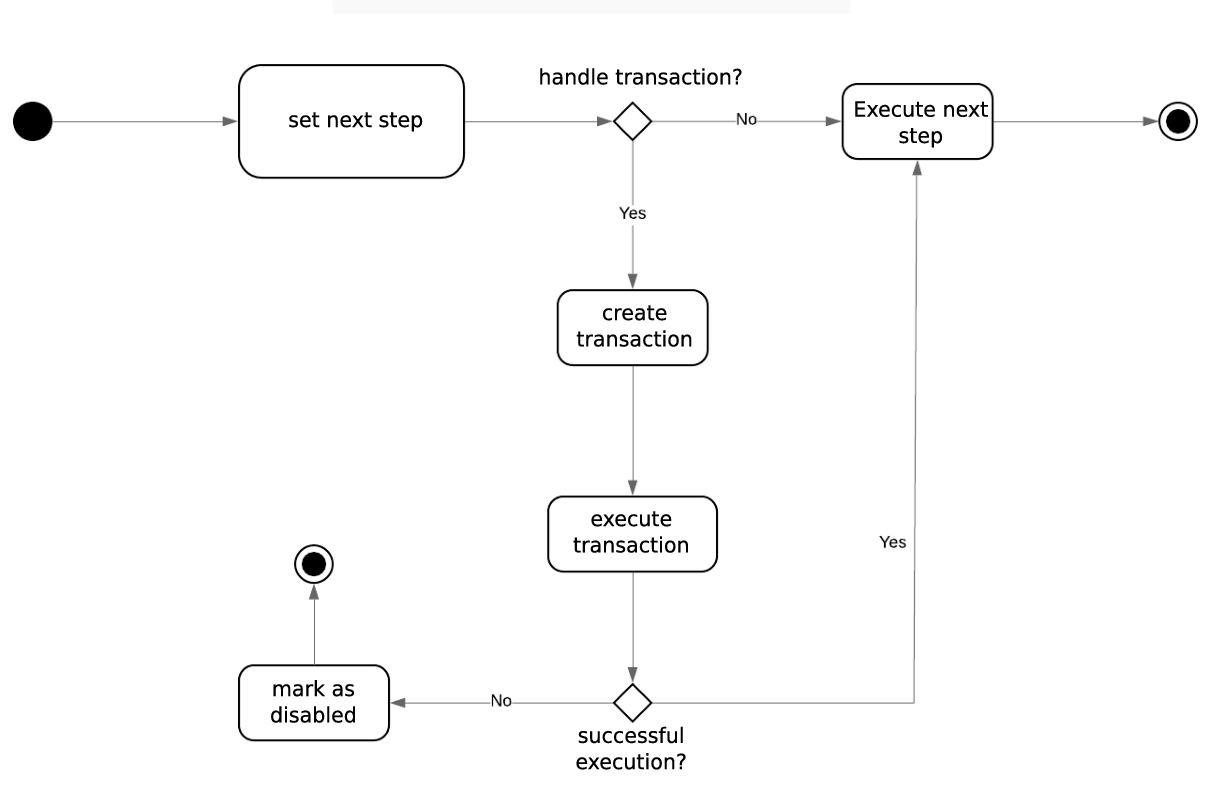}
\caption[diagrama-secuencia-simple]{Transition loop}
\label{fig:transition-loop}
\end{figure}

The main feature of the proposed framework%
\footnote{The source code for the framework developed is available at \url{https://github.com/GRISE-UPM/ml_server_rest}.} %
is the \textit{transition loop} shown%
\footnote{This figure incorporates transaction management, which we will discuss later.} %
in Figure~\ref{fig:transition-loop}. A choreography can be described as a state transition diagram. The transition loop determines which transition should be executed in each case, performs it, and upon completion assigns the corresponding state to the ubiquitous device.

\begin{figure}[htbp!]
\centering
\includegraphics[width=7cm]
{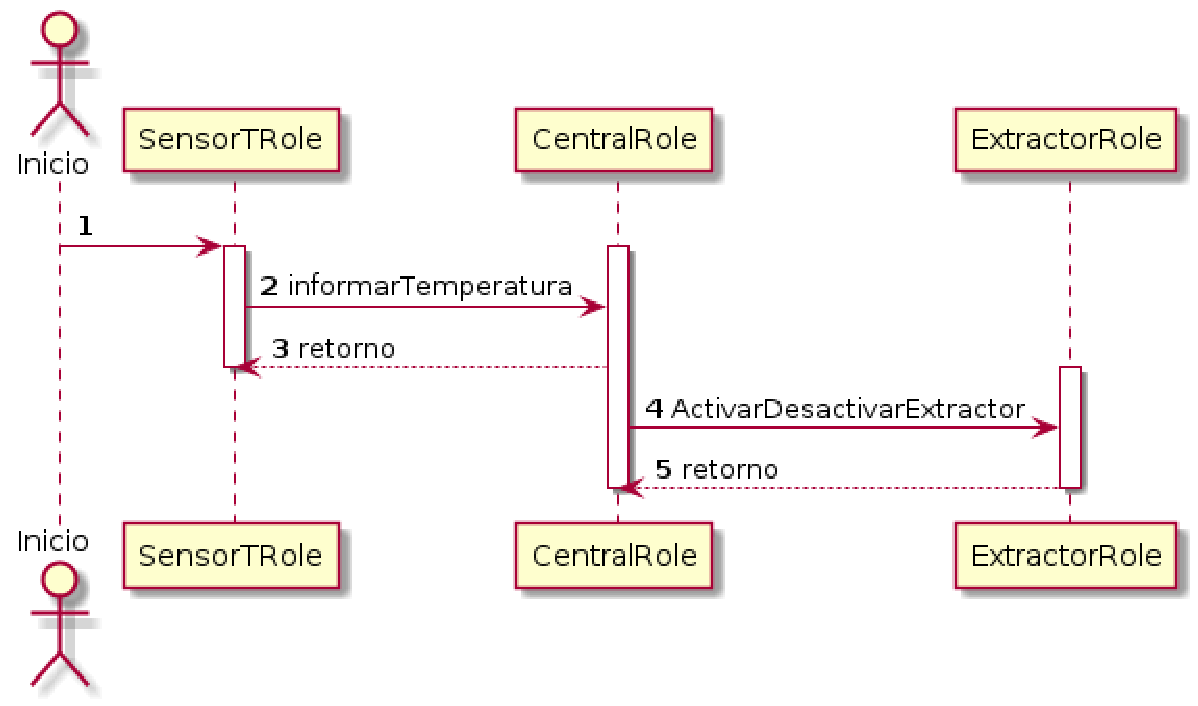}
\caption[diagrama-ejecucion-siguiente]{Sequence diagram showing the choreography proposed in the example.}
\label{fig:coreography-example}
\end{figure}

For example, consider the following choreography: A device named \texttt{SensorTRole} produces a communication with another device, \texttt{CentralRole}, which in turn relates to another device, \texttt{ExtractorRole}. In this case, \texttt{SensorRole} sends a message to \texttt{CentralRole} reporting the ambient temperature. Based on this temperature, \texttt{CentralRole} decides whether to turn on an extractor to lower the temperature. If affirmative, it sends a message to activate the extractor to \texttt{ExtractorRole}.  

This choreography can be graphically represented by the sequence diagram in Figure~\ref{fig:coreography-example} and formally described using WS-CDL in Listing~\ref{fig:listing1}. The \textit{transition loop} is the execution engine of the choreography, and therefore, according to the Role being executed, it determines the next step or interaction to perform. Before doing so, it checks for any transactions to implement or continue executing, and then it completes its participation in the choreography.

\begin{lstlisting}[language=xml, caption={Description of the choreography in WS-CDL. Only a piece of the entire XML specification is shown}, label={fig:listing1}]
<interaction name="tempAction" operation="tempReport">
	<participate relationshipType="tns:SensorT_Central" fromRole="tns:SensorTRole" toRole="tns:CentralRole"/>
	<exchange action="request" name="tempReport" informationType="tns:tempInfoType">
		<send variable="cdl:getVariable(tns:tempData,SensorTRole)"/>
		<receive variable="cdl:getVariable(tns:tempData,CentralRole)"/>
	</exchange>
</interaction>
<interaction name="extractorAction" operation="extractorED">
	<participate relationshipType="tns:Central_Extractor" fromRole="tns:CentralRole" toRole="tns:ExtractorRole"/>
	<exchange action="request" name="extractorED" informationType="tns:ActionType">
		<send variable="cdl:getVariable(tns:Action,CentralRole)"/>
		<receive variable="cdl:getVariable(tns:Action,ExtractorRole)"/>
	</exchange>
</interaction>
\end{lstlisting}

\subsubsection{Implementation}

This choreography can be easily implemented using the proposed framework. We used PHP and C++ as programming languages. These languages were selected based on the devices used for the concept test (Arduino, Raspberry Pi, and PC), *pero cualquier otro lenguaje de programación podría haber sido utilizado.*  

In the case of the example choreography (Listing~\ref{fig:listing2}), a class is implemented in C++ (Listing~\ref{fig:listing2}) that only needs to encode specific methods to capture and process messages, but at a high level of abstraction, as the framework's classes handle the lower-level details.

Listing~\ref{fig:listing2} is a template of a class that implements and uses the framework. In this case, the code that implements the \texttt{CentralRole} device is shown, which is part of the interactions in the previous definition. The roles \texttt{SensorT} and \texttt{Extractor} have the same simple implementation as \texttt{CentralRole}. 
 
\begin{lstlisting}[language=c++, caption={Implementation using the framework}, label={fig:listing2}]
#include "ClaseX.h"
#include <connection.h>
#include <RESTWebServer.h>

CentralRole::CentralRole() : choreographyService(){
    chor_act_role = "CentralRole";

    return;
}
int CentralRole::start(){
	return choreographyService::start();
}
void CentralRole::execute(){
	if (strcmp(method,"tempReport") == 0){
		if (this->_verb != 'O'){
			pServer->HTTPresponse("405 Method not allowed","");
			return;
		}
		String response = this->tempReport(method,tempData);
		strcpy(sResponse,"{\"result\":\"");
		strcat(sResponse,respuesta.c_str());
		strcat(sResponse,"\",\"token\":");
		strcat(sResponse,_token);
		strcat(sResponse,"}");
		pServer->HTTPResponse("200 OK",sResponse);
		choreographyService::execute();
}
}
\end{lstlisting}

\subsubsection{Class hierarchy}

The code in Listing~\ref{fig:listing2} uses a set of classes provided by the framework. Among these classes is the \texttt{choreographyService} class, which manages the execution of the choreography. This includes determining which interactions should be executed next, which messages should be sent, and managing the data received from the messages invoking this device.  

For the example choreography in Listing~\ref{fig:listing1}, the \texttt{choreography} class is responsible for receiving and processing the incoming messages and determining which devices to invoke based on the interactions defined in the choreography.  

We must extend the \texttt{choreographyService} class and start coding the specific behavior of the role in the \texttt{ejecutar} method.

\begin{figure*}[]
\begin{center}
\includegraphics[width=0.7\linewidth]{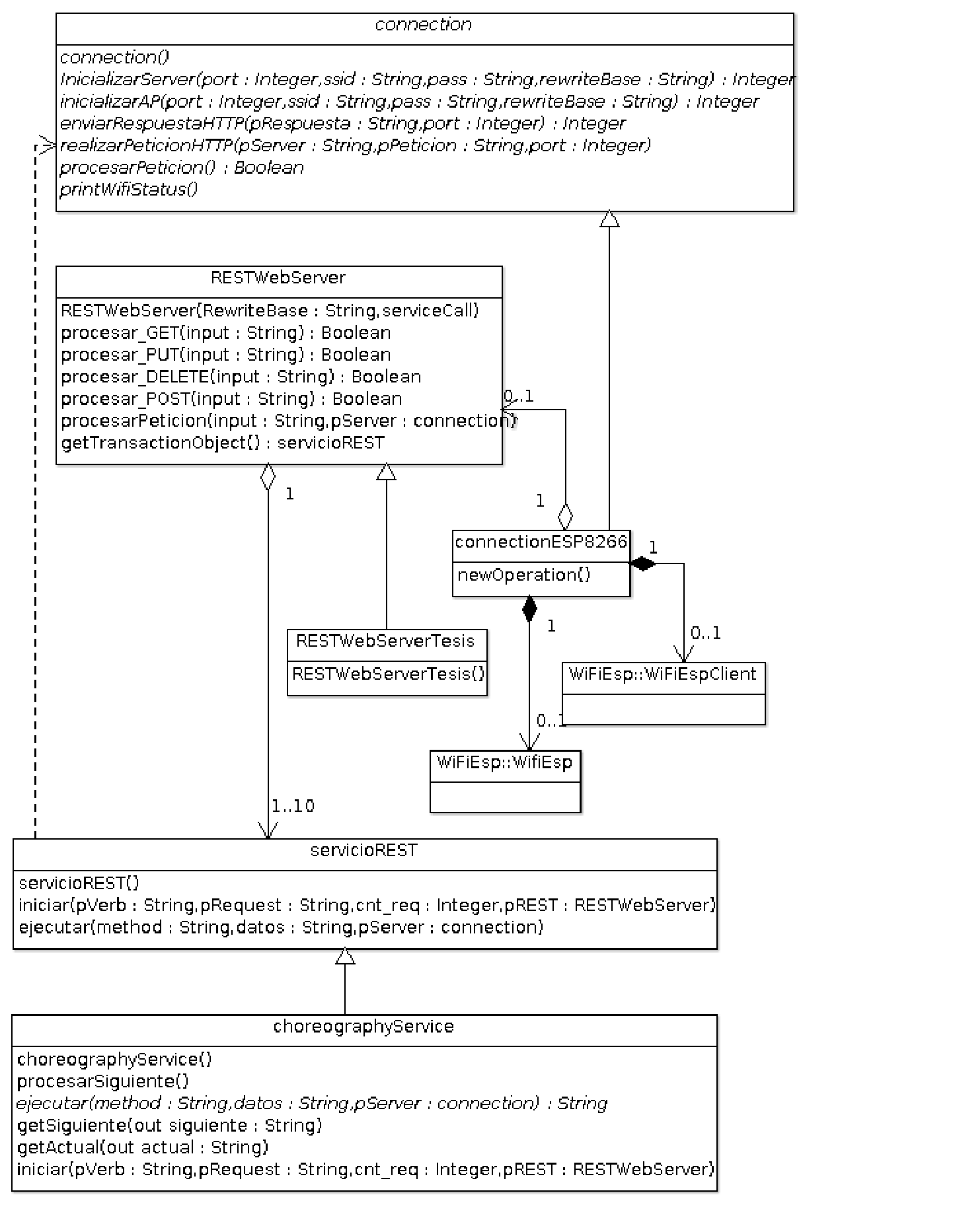}
\caption[diagrama-clases]{Class Diagram}
\label{fig:listing3}
\end{center}
\end{figure*}

\subsection{Particular characteristics of pervasive systems}

Table \ref{t1_5} outlines the limitations of ubiquitous devices \cite[p.~19-76]{ubiquitousPoslandStefan2009}. Such limitations must be taken into account by the proposed framework to be effectively used in practice. Below, we describe how we have addressed each of these limitations.

\begin{table*}
\centering
\caption{Features of ubiquitous devices}
\label{t1_5}

\begin{tabular}{p{0.18\textwidth}p{0.36\textwidth}p{0.36\textwidth}} 

\hline \hline
Feature & Associated Problem & Proposed Solution Alternative\\
\hline \hline

Limited resources & These limitations imply that the device may run out of battery while in the middle of executing a choreography, or may not have enough memory or computational capacity to run the software \cite[p.~19,76]{ubiquitousPoslandStefan2009}. & \multirow{2}{0.36\textwidth}{Implement ad-hoc solutions that limit resource usage, or use lightweight services.}\\ \cline{1-2}

Invisibility & Many ubiquitous devices try to remain unnoticed, e.g., wearable devices. Being so small, they have significant resource limitations \cite[p.~48]{ubiquitousPoslandStefan2009}.\\ \hline

Heterogeneity & Devices have proprietary APIs, which makes their interoperability difficult \cite[p.~19,76]{ubiquitousPoslandStefan2009} \cite[p.~39,43,44]{ubiquitousKrumm2010}. & Use REST services, which are standard, open, and fit perfectly within SOA and WSCDL.\\ \hline

Connectivity & Communication between devices can be done in different ways, e.g., Bluetooth, WiFi, or even wired network. & Use standards, e.g., TCP/IP, to connect the devices to each other.
\\ \hline

Embedded & Some devices, e.g., wrist bands, do not connect to any other device, but to a coordinator device, e.g., a mobile phone, usually of higher capacity \cite[p.~4]{ubiquitousPoslandStefan2009}. & This is a type of restricted connectivity that has no implications from the WSCDL choreography point of view.\\ \hline

Autonomy & Devices select the connectivity that best suits their needs \cite[p.~15]{ubiquitousPoslandStefan2009} in order to move from one place to another, or even disappear under certain circumstances, e.g., stealth mode. This implies that the rest of the members stop receiving a response from this device \cite[p.~76]{ubiquitousPoslandStefan2009} \cite[p.~41]{ubiquitousKrumm2010}. & \multirow{2}{0.36\textwidth}{Implement mechanisms to compensate for the transient (or permanent) disappearance of devices. There are multiple solution alternatives. We will apply two strategies:

\begin{itemize}
\item Apply time-outs, as specified in the WSCDL specification, and perform a rollback of the actions taken during the choreography.
\item Make extensions to the WSCDL specification to allow the use of replacement devices or digital twins.
\end{itemize}} \\ \\ \\ \\ \\ \\ \hline

Security and privacy & Communications between devices must be authenticated and under encrypted transport layers \cite[p.~49]{ubiquitousPoslandStefan2009}. & Implement the upper layers of SOA within the devices (see Figure~\ref{f2_5}). \\ \hline \hline

\end{tabular}
\end{table*}

\subsubsection{Limited resources and invisibility}\label{subsec:recursoslimitados}

Ubiquitous devices are based on commercial microcontrollers. The most common controllers use architectures ranging from 8 to 32 bits, with program memory (FLASH) between 4KB and 2MB and main memory (SRAM) between 2KB and 1MB. Given that ubiquitous devices do not have an operating system and the programs are simple (for example, they lack a graphical interface, or provide one using HTML), the upper limits of architecture/FLASH/SRAM do not present significant limitations to the programmer. However, attention must be paid to the lower limits, where clear limitations do exist.

\textbf{Program memory} -- Program memory prevents the use of commercial web servers like \textit{apache}%
\footnote{Additionally, we must consider that ubiquitous devices do not have dynamic libraries; statically compiling web servers increases the memory size required even further.}%
. The \textit{lighthttpd} server could be used, as the object code (without additional libraries) is approximately 256KB. Even so, this is a considerable size for many microcontrollers.

In our case, we chose to develop a simple web server, where we focused on handling requests based on Representational State Transfer (REST). This simplifies the task since we only implement a REST API that supports the four main REST verbs: GET, POST, PUT, and DELETE. Therefore, the class designed for this purpose, as shown in Figure \ref{fig:listing3}, implements these verbs and handles the requests. This class is extremely simple, and the SRAM memory it consumes is on the order of 50 bytes.

\textbf{Main memory} -- Main memory is a serious problem. The program variables, data, and information to be transferred between devices can be considerable. However, the main issue arises with the specification of the choreography in XML format, as describing it in XML consumes a lot of memory to store, read, and execute it at runtime. Therefore, its inclusion in ubiquitous devices becomes impractical. As an example, the size of the choreography used for the proof of concept in this work is 8 KB, which is the total SRAM memory available on an Arduino Mega 2560 board (like the ones used in this work). In Section~\ref{subsection:Execution}, where we work on the Proof of Concept, we see how it performed on the devices.

To solve this limitation, the choreography specification has been reduced to what concerns each device, so that only the interactions that are the responsibility of the corresponding role are loaded into memory. In Listing~\ref{fig:listing2}, the device fulfilling the \texttt{CentralRole} only has in memory the interactions with \texttt{ExtractorRole}, and the other parts of the choreography specification are not kept in memory.

\textbf{CPU} -- The CPU used varies depending on the manufacturer, but one of the most popular options is ARM chips. The simplest chip ARM currently produces is the Cortex-M0+, which has a score%
\footnote{\url{https://developer.arm.com/Processors/Cortex-M0-Plus}} %
of 2.36 CoreMark/MHz. To put this value in context, the Pentium processors evaluated%
\footnote{\url{https://www.eembc.org/coremark/scores.php}} %
by the EDN Embedded Microprocessor Benchmark Consortium have scores ranging from 1.67 to 3.73 CoreMark/MHz. Consequently, the computing power of microcontrollers does not pose a limitation, as we have empirically verified in the proof of concept (see Section~\ref{sec:proof-of-concept}).

\textbf{Battery} -- In the various tests conducted, the devices were connected to a continuous power source. While battery life estimates fundamentally depend on the size and power of the battery, we can estimate that, according to the design used for the operation of the devices \footnote{API calls to the microcontroller are used to put the Arduino board into sleep mode, so it only becomes active upon receiving a request via the serial port through the used Wi-Fi module.} and the number of requests received within a given time, the battery life could exceed 5 days with a demand of 5 requests per hour.

\subsubsection{Heterogeneity}

The integration of different elements, devices, or appliances that have the ability to connect with other elements is not straightforward, as they use proprietary APIs. For example, if we wanted to integrate a dishwasher and a washing machine into a device choreography to perform a specific task, we would need to create an account in Amazon's Home Connect, as well as link an account with an Echo device, also from Amazon. Then, from the Alexa app, we could use some available functions. In some cases, such as with Google, it is necessary to have a paid account to use these services. In other words, not only is the software proprietary, but there may also be associated costs.

The framework proposed in this work makes use of the already standardized REST verbs (GET, POST, and PUT), which are already integrated into the proposed code. This only requires the coding of specific methods related to the services offered by the device. In Listing~\ref{fig:listing2}, this can be observed, where the \texttt{execute} method is coded. The interaction of the different devices is established through the specification of the choreography, written in the standardized WS-CDL language, as can be seen in Listing~\ref{fig:listing1}.

\subsubsection{Connectivity}

We conducted pilot tests using WiFi and Bluetooth connections. The experience gained, contrary to what the literature suggests, indicates that connectivity does not pose a major issue. The most common third-party communication microcontrollers, e.g., the Esp8266 Esp 01 Wifi Module, come with embedded communication software and provide an API for use. For example, the aforementioned microcontroller uses built-in AT commands, which were accessed via a high-level library (WiFiEsp) \footnote{https://docs.arduino.cc/libraries/wifiesp/}. This greatly simplifies application development, as in the case of the present framework, and does not introduce additional resource consumption that would compromise the framework’s execution.

Connectivity could pose a problem in low-capacity ubiquitous devices with more basic in-board communication microcontrollers. In this case, it might be necessary to develop a substantial amount of code for network communication, leaving fewer resources available to run the framework, e.g., consider the code required to implement the I2C protocol \cite{i2c:2021}. However, embedded communication microcontrollers do not seem to be common, as communication standards evolve rapidly, complicating the design of boards. Manufacturers prefer to use third-party communication microcontrollers.

\subsubsection{Embedded}

As with connectivity, the fact that ubiquitous devices are embedded in a larger system does not introduce any complexity, as long as: (1) the communication protocols used are standardized, e.g., Bluetooth, and the hardware resources are not too limited, e.g., having a minimum SRAM of 8 KB, since otherwise the framework could not be executed.

\subsubsection{Autonomy}

So far, the application of SOA concepts to ubiquitous systems has been relatively straightforward. Although there are certain limitations (memory, battery, etc.), ubiquitous devices can implement SOA using specific software and data structures. However, autonomy is the characteristic that most differentiates traditional software systems from ubiquitous systems, and it is where SOA concepts will need to be adapted.

Claro, aquí tienes la traducción respetando los tags de LaTeX y lo que ya está en inglés:

\textbf{Mobility} -- One of the most immediate effects of the mobility of the devices is that, for a certain time, they stop being available to participate in a choreography. The disappearance can take place at any moment (before or after receiving messages), and its duration can be short (for example, a temporary connectivity issue), or long (for example, due to a network change that involves a relatively long data propagation time). Bottlenecks in processing that result in response times that, in practice, are equivalent to a disappearance cannot be ruled out either.

This type of problem also occurs in SOA systems and, consequently, the WS-CDL specification includes time-out mechanisms for the previous cases, which we have implemented in the choreography. In this case, the framework, after a time-out issue occurs, marks the specific execution of the choreography as unusable within a given time range, so any execution that has been interrupted and attempts to invoke it will be automatically rejected.

To demonstrate that the SOA protocol stack shown in Figure~\ref{f2_5} is applicable to ubiquitous systems, we have made a partial implementation of the WS-Coordination specification \cite{OASIS:WSC:2007}. This implementation provides the possibility for the choreography to carry out an atomic distributed transaction through the use of fully distributed and controlled commit and rollback, allowing the developer not to focus on coding this feature, as by simply instantiating certain objects, the framework will handle the transaction transparently. The limitation of this implementation is that the WS-Coordination specification is based on the SOAP protocol, and in our case, we use REST as the web service protocol, so its application was mainly based on the model presented by Guy Pardon et al. \cite{10.1145/2567948.2579221}.

\textbf{Inaccessibility due to addressing issues} -- One aspect we have not addressed in the framework is connectivity loss due to network changes, e.g., IP and domain changes, which, while not exclusive to ubiquitous devices, becomes more evident in these due to their mobility. There are works that address this dynamic discovery topic, which could be applied in the case of disappearances or inaccessibility. For example, the work by Najar et al.\cite{Najar2014421} proposes a dynamic discovery mechanism, which can be used for both choreographies and orchestrations. Another example is the work by Palmieri \cite{Palmieri2013693}, where the issue of device composition is approached from the point of view of dynamic discovery of these devices. Aura \cite{auraproject} and Gaia \cite{gaiaproject} are examples of frameworks that have dynamic discovery, but in a centralized manner as if it were an orchestration.

The simplest solution, and one that is used in many IoT devices, consists of offering a centralized device registration mechanism\footnote{As is the case with the environments and frameworks mentioned in \ref{sec:background}}. The centralized registration could be easily implemented in our framework, but this solution introduces an indirect orchestration mechanism and, therefore, a single point of failure. For this reason, we have not included this functionality in the framework.

On the other hand, inaccessibility due to addressing issues arises from mobilities that we believe are not common (think, for example, about the frequency with which a temperature sensor moves between physical locations without requiring reconfiguration). It is, therefore, a problem that occurs in very specific applications and could be implemented through an \textit{ad hoc} centralized registry.

\textbf{Clones vs. exceptions} -- In both of the previous cases, the developer must plan exceptional processing mechanisms that generally involve aborting the choreography and, if necessary, applying a \textit{rollback} mechanism. An alternative that partially addresses the disappearance of certain ubiquitous devices through \textit{clones} or \textit{digital twins} \cite{jiang2021industrial}\cite{digitaltwinMatlab}.

Clones are replicas of devices that have disappeared, which could reside on another device or even in a cloud computing service such as Amazon's AWS-IoT. The processing that was the responsibility of the disappeared device would be delegated to the clone. This can be useful in various scenarios: ubiquitous devices that do not interact with the environment, or do so in predictable or low-risk situations, e.g., an environmental temperature sensor with a history, a display providing information to the user, etc.

Using WS-CDL, clones can be defined in several different ways:

\begin{itemize}
\item Define it as a role: in this case, an extension of the specification language should be made by adding attributes to specify whether the role is the primary one or not. This option would not fully reflect the reality of the device, as the role is already defined by the main device carrying out the choreography. While technically and theoretically, it would not be correct when reading and interpreting the choreography in the coding process.
\item Define it as a participant: this would be the same case as the previous point, or we would have to define a new type of participant, which would not accurately reflect what happens in the choreography, as it is not a new participant, but rather a cloned one. Again, the problem is not with the coding itself, but it would not be technically or theoretically correct.
\item Inline replacement: through this alternative, we would replace the device at runtime in a way that is completely transparent to the choreography and its specification. For example, AWS IoT services provided by Amazon could be used, or other alternatives, or an \textit{ad hoc} implementation for this case. This option is more complex to implement and may involve implementation costs, though it has the advantage that no changes are needed in the specification. However, it might obscure an interpretation that is not always desirable, since if it is not properly documented, or if the paid service fails, the causes of the cancellation or non-execution of the full choreography cannot be determined.
\item Define a new \texttt{clone} element: this would involve adding to the WS-CDL specification, including a section for defining the devices that will act as replacements for the original or true roles within the choreography. In this case, it will be necessary to define which role and participant it would be replacing and, if needed, other attributes to define it as best as possible, such as the interface it implements, the type (on-demand or permanent), etc.
\end{itemize}

The best applicable solution is to use the \texttt{clone} element. This is an alternative that meets all the transparency, technical, and theoretical requirements both at the choreography level and its execution. We have added it to the specification as follows:

\begin{verbatim}

<cloneType name="NCName" 
           interface="QName"?
           type="on-demand"|"permanent">
           <roleType typeRef="QName"/> +
</cloneType>

\end{verbatim}

The meaning of the attributes is as follows:

\begin{itemize}

\item The \texttt{interface} attribute is defined as optional, identifying the WSDL interface.

\item The \texttt{name} attribute is used to specify a different name for each \texttt{cloneType} element declared within a choreography package.

\item The \texttt{type} attribute is used to specify the type of usage for the \texttt{cloneType} element, with the possible values being:

    \begin{itemize}
    
		\item \texttt{on-demand} - The device that will act as a clone will only be used in a specific case of a particular relationship.
    
		\item \texttt{permanent} - Once the device is activated as a replacement, it remains visible to the rest of the choreography. This is the default behavior.
    
    \end{itemize}

\end{itemize}

Within the \texttt{cloneType} element, one or more \texttt{roleType} elements identify the \texttt{roleTypes} that must be implemented by this \texttt{cloneType}. Each \texttt{roleType} is specified by a \texttt{typeRef} attribute of the \texttt{roleType} element, which is defined earlier in the choreography declaration. The \texttt{QName} value of the \texttt{typeRef} attribute in the \texttt{roleType} element must reference the name of a \texttt{roleType} declared previously.

\textbf{Inaccessibility due to communication or battery issues} -- It is clear that no solution can be provided through software.

\subsubsection{Security and privacy}

As we have seen in Figure \ref{f2_5}, there are different layers in the SOA specification corresponding to security and privacy (WS-Security, WS-Policy) \url{https://docs.oasis-open.org/wss-m/wss/v1.1.1/os/wss-SOAPMessageSecurity-v1.1.1-os.html}. These layers, defined and specified for the SOAP protocol, can be implemented on the REST protocol used in this framework, even in a more simplified manner.

The implementation of these security layers implies an overhead on request processing that significantly affects the response time of the ubiquitous devices used. Additionally, this encoding would consume main memory, which should be thoroughly evaluated due to the limitations already mentioned.  
Currently, we are researching and evaluating alternatives for implementation in the next prototype cycle, as it has not been implemented in this version of the framework and the proof of concept presented here.

\section{proof-of-concept}\label{sec:proof-of-concept}

We have selected a domain related to vehicle traffic problems along highways, routes, or roads. Vehicles that travel along a road communicate with beacons distributed along the track. This communication can be bi-directional when the vehicle reports a problem or the beacon gives notice of possible problems along the road. In turn, beacons can communicate with emergency or coordination stations as appropriate. 

\begin{framed}
A bus with passengers moves from one city to another. The bus is equipped with a sensor that detects the possibility that the bus driver suffers a heart attack. From that moment on, the vehicle warns the driver (either visually or by audio signal) and, in turn, tries to communicate with a beacon on the road so that help arrives as soon as possible. The beacon also communicates with other beacons, triggering alerts to nearby vehicles so that they can take preventive action.
\end{framed}

\subsection{Devices}

\begin{figure}[htbp!]
\centering
\includegraphics[width=6.5cm]
{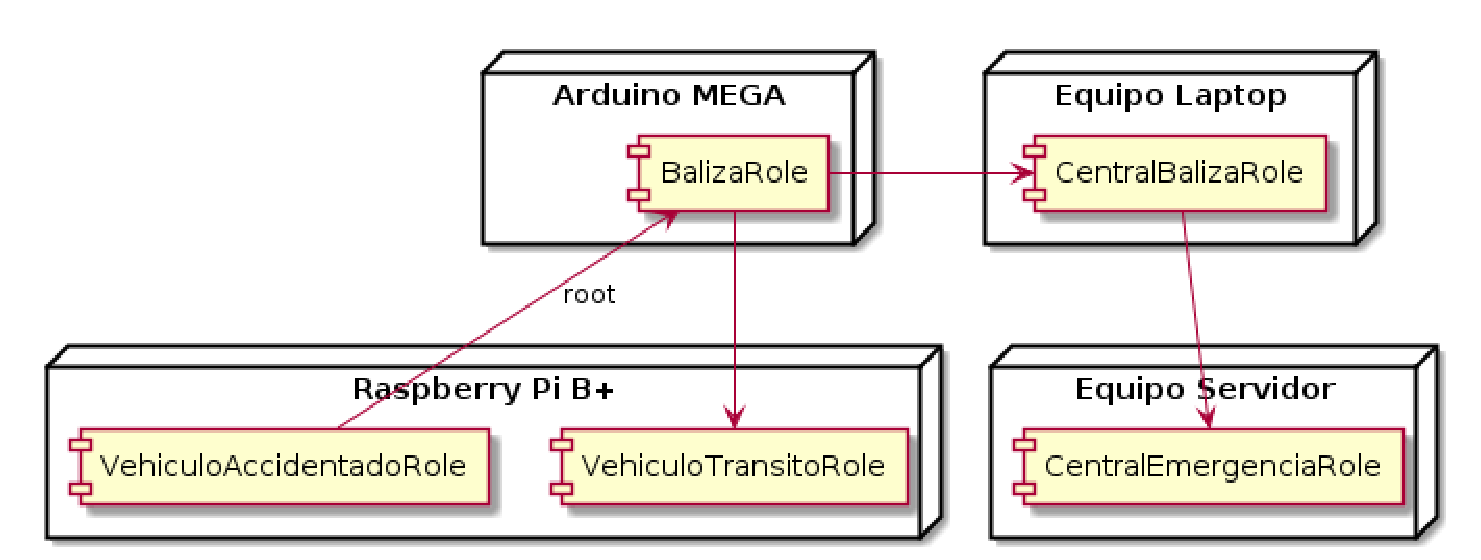}
\caption[diagrama-despliegue-coreografia]{Devices and Roles Display Diagram}
\label{f3_6}
\end{figure}

The proof-of-concept architecture can be seen in Figure \ref{f3_6}, where each device is seen with the role it plays within the choreography. The Arduino board represents the role of a \texttt{Baliza}, the RaspberryPi implements the devices with the roles of \texttt{VehiculoAccidentado} and \texttt{VehiculoTransito}, and the equipment with more processing capacity play the roles of \texttt{CentralBaliza} and \texttt{CentralEmergencia}. The devices used for the construction of the concept test are the following:

\begin{itemize}
	\item Server computer: Executing Central Vehiculo Accidentado Role. 
	\item Laptop computer: Notebook connected to a VPN with the main server mentioned above. Executing Central Emergencias Role.
	\item RaspberryPi B+: Lighttpd web server version 1.4.x, PHP version 5.5.x working in CGI mode. This computer is connected to the VPN with the server equipment mentioned above. Executing Central Balizas Role.
	\item Arduino Mega 2560 development board. Executing Baliza Role.
	\item Arduino Nano V3 development board, of only 2K memory available for data and a 16-bit processor. Executing Vehiculo Role.
\end{itemize}

\subsection{Device limitations}

Next, we will address each of the limitations of the devices used based on the limitations presented in section \ref{sec:framework}.

\subsubsection{Limited resources and invisibility}
We can highlight the following devices:
\begin{itemize}
\item{Arduino:} It has memory limitations (both program and main memory), CPU limitations, and battery dependency. The basic characteristics are: 8k memory for data and a 16-bit processor.
\item{Raspberry Pi:} We have also determined that these devices have fairly limited processing capacity in terms of speed and capacity, although connectivity is very good. The basic characteristics are: 32-bit single-core processor, 1GB RAM.
\end{itemize}

\subsubsection{Heterogeneity}
Heterogeneity is represented in the devices used, as they belong to different providers and have different technical characteristics.
\begin{itemize}
\item{Arduino}: It does not have an operating system, only a bootloader responsible for launching the program execution. Connectivity is achieved through the WiFi module (ESP8266-01).
\item{Raspberry Pi}: It has the Raspbian operating system, and connectivity is provided by an integrated Ethernet board.
\item{Server}: PC with 16GB RAM, 4-core processor connected to a public access IP address, Linux operating system, Apache 2.4.18, PHP version 7.0.15, and a PostgreSQL 9.3 database.
\item{Notebook}: Laptop with 12GB RAM, dual-core processor running similar characteristics to the Server.
\end{itemize}

\subsubsection{Connectivity}
The only device used with limited connectivity has been the Arduino boards, whose connectivity depends on external modules that do not always have good connection stability. Specifically, we used an ESP8266-01 module, which proved suitable for this proof of concept, with some communication issues due to the inherent instability of the module's firmware.

\subsubsection{Embedded}
The two devices used in this proof of concept that meet this characteristic are the Arduino and the Raspberry Pi, as they are installed in a vehicle or in a beacon on the road, being invisible to humans.

\subsubsection{Autonomy}
In this section, the two devices with limited autonomy \footnote{The autonomy time depends on the device; in the case of the Raspberry Pi, we are talking about approximately 4 to 5 hours, while for the Arduino, depending on the type of battery, it can exceed 24 hours.} are as follows: 
\begin{itemize}
\item{Arduino:} This device has autonomy dependent on the type of available battery to which it is connected. For this proof of concept, however, this device was programmed to enter hibernation mode after fulfilling the requested tasks, thereby extending the autonomy dependent on the battery. 
\item{Raspberry Pi:} This device can operate connected directly to electrical power through a transformer or to a battery. In this case, the autonomy is much shorter than that of the previous device because it cannot be set to hibernation mode, in addition to its higher energy consumption compared to the Arduino board.
\end{itemize}

\subsubsection{Clones}

In our case, we will use this \texttt{cloneType} element to define a device that will replace the role \texttt{VahiculoTransitoRole} in case it cannot be contacted on the road at the time of the accident, and it will be notified when it can reconnect. To achieve this, we will define the following \texttt{cloneType}:

\begin{verbatim}

<cloneType name="VehiculoTransitoClone" 
 type="permanent">
 <roleType typeRef="VehiculoTransitoRole"/> 
</cloneType>

\end{verbatim}

Where the new element is named \texttt{BalizaClone} and will replace the role \texttt{BalizaRole} in case of its disappearance.

\subsection{Conectivity}

There are currently various projects studying V2X connectivity (vehicle to any other device or entity), which includes connecting a vehicle to other vehicles, pedestrians, traffic signals, road signs, infrastructures of all kinds, etc. The vehicle can communicate with the outside world through its built-in sensors and wireless connectivity. Two wireless communication technologies stand out \cite{dsrcversus4g2017}: Dedicated Short-Range Communication (DSRC) and Cellular Vehicle-to-Everything (C-V2X). DSRC is a modified version of the Wireless Local Area Network (WLAN) protocol, IEEE802.11p, whereas C-V2X utilizes cellular radio instead of WLAN.

Given the similarity between the 802.11 standards for both LAN and DSRC, we adopted the 802.11g/n protocol for the proof of concept. This additionally has the advantage that 802.11g/n communication modules for Raspberry and Arduino devices are easy and inexpensive to obtain.

\subsection{Development}

The choreography in WS-CDL language used for this proof of concept can be found in Annex~\ref{sec:anexos}, where, due to space constraints, only the interactions carried out are shown in Listing~\ref{fig:coreografia_accidente_wscdl}.

To execute the choreography, a series of classes had to be coded in both PHP and C++ to enable its execution. While the source code is relevant in both languages, the most complex development was for the Arduino board, as the storage capacity for the code is limited to 256 KB, and the data usage during execution is limited to 8 KB. The code occupies approximately 32 KB, which is only 13\% of the total available space, and uses 3 KB of SRAM memory, representing 35\% of the total available for variables and local data. 
Since these boards lack the ability to host a web server due to the absence of an operating system, all the necessary components to process REST requests and their corresponding responses, as well as to interpret and execute the relevant portion of the choreography, had to be developed in C++.

\subsection{Execution}\label{subsection:Execution}
During the execution of the choreography, analyses of the memory usage were conducted, focusing primarily on dynamic memory or heap, as this is used by the code to store various local variables and instantiated code objects at any given moment. From this analysis, it was observed that during certain points in the choreography execution by a device, the usage of dynamic memory (SRAM) reached 6 KB. In Section~\ref{subsec:recursoslimitados}, we address these characteristics in greater detail.

\subsubsection{Reducing memory usage}
Within the efforts made to incorporate the choreography definition into ubiquitous devices, we explored the possibility of using a library to include the choreography definition in a compressed format on each device. To this end, we analyzed various alternatives, all based on the Efficient XML Interchange (EXI) standard for representing XML documents \cite{exip-spec:2014}. Ultimately, we selected the EXIP library \cite{exip-library}, developed in the C programming language, with its primary goal being usage in network-embedded devices.

The implementation of EXIP in our framework, particularly on Arduino development boards, proved unsuccessful because the memory required to execute even a small test far exceeded the SRAM capacity of an Arduino Mega 2560 board, which has one of the largest SRAM capacities available. Although it was possible to generate an Arduino library for the EXIP specification that occupies only 18\% of the code storage space available on this board, it could not be used because the required SRAM memory was 9 KB, exceeding the allowed limit (113\%). Additionally, the choreography in compressed format occupies approximately 5 KB \footnote{The uncompressed XML format choreography occupies about 8 KB.}.

The objects instantiated from the framework with the highest SRAM usage are those that handle REST requests, although they do not saturate the maximum available memory capacity during execution. These classes primarily manage communication and the parsing of requests, consuming about 1 KB of SRAM during execution.

\subsubsection{Processing capacity}

The various executions demonstrated an execution time of approximately 200 milliseconds, as indicated in the graph associated with the execution of the choreography shown in Figure~\ref{f1_6}, where the participating devices, their IP addresses, and the date-time of execution are displayed. Over 400 executions of the proof of concept, the average execution time for the choreography was 3.42 seconds, including the handling of distributed transactions.

\begin{figure}[htbp!]
\centering
\includegraphics[width=9cm]
{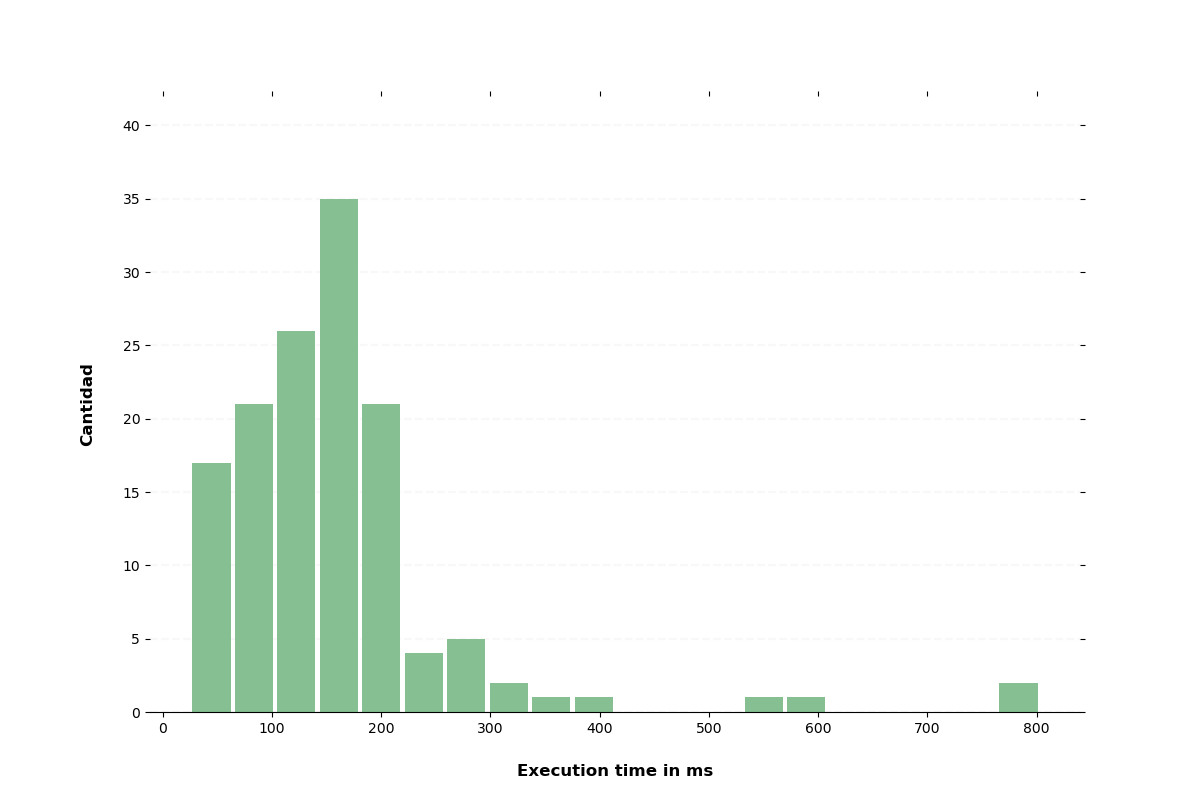}
\caption[Histograma de tiempos de ejecución de la coreografía en milisegundos]{Histogram}
\label{histograma}
\end{figure}

In Figure~\ref{histograma}, we can observe a histogram of execution times across more than 400 runs \footnote{Of those 400 executions, those with very wide dispersion, due to errors, were discarded.}.

\subsubsection{Choreography visualization}

\begin{figure*}[htbp!]
\centering
\includegraphics[width=\textwidth]{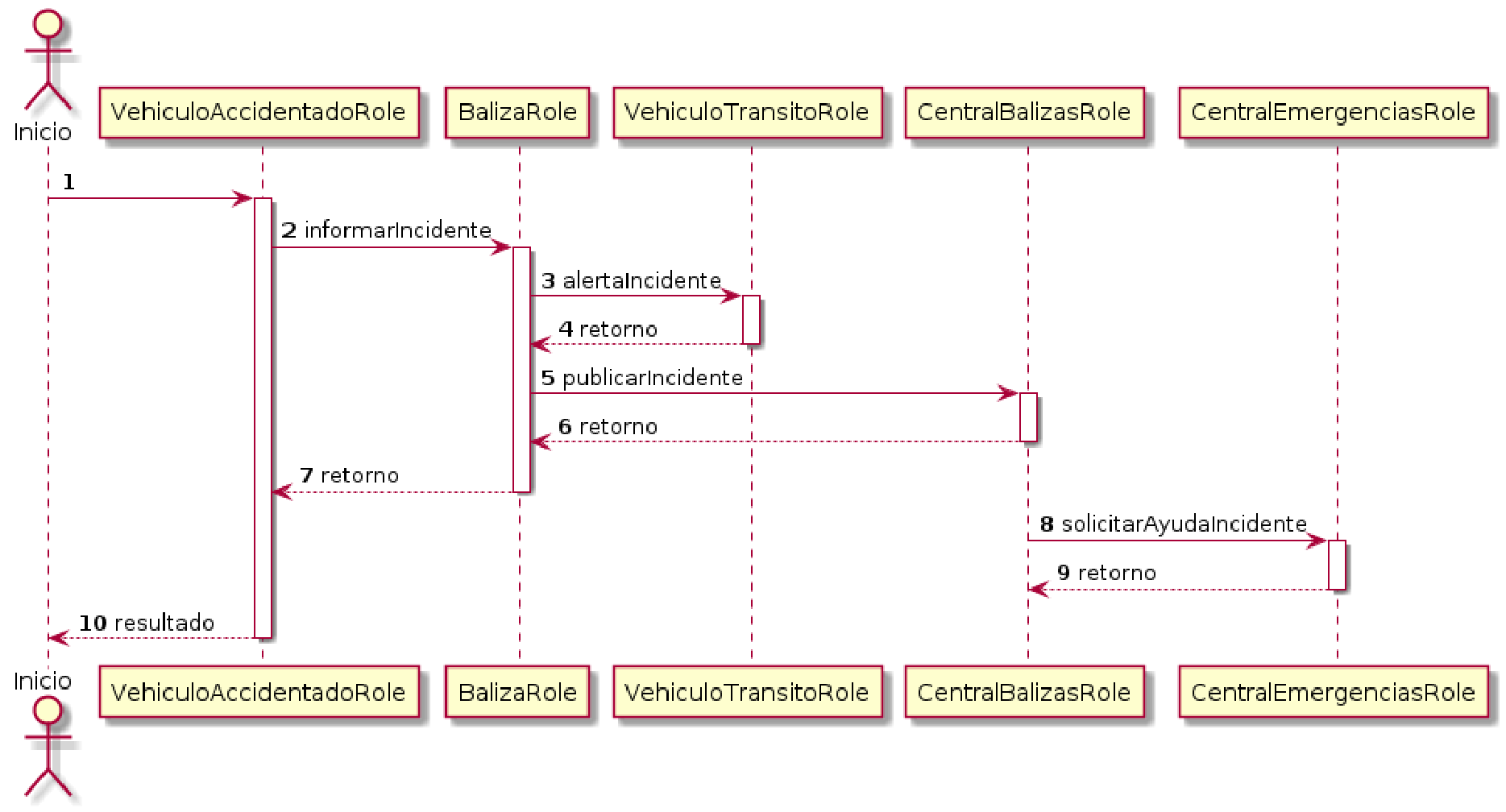}
\caption{Choreography execution sequence diagram}
\label{f1_6}
\end{figure*}

The sequence diagram of Figure \ref{f1_6} represents the execution of the choreography. This figure is automatically generated by the framework for evaluation purposes. This graphing tool is available on the public server used for the various tests. To access it, you must go to the following address \url{http://170.210.127.69/tesis/graficar_coreografia.php}, where you need to enter a token code that identifies the execution of the choreography. After entering the code, the program displays the graph. In the graph, we can visualize the requests made by the different roles/devices executing the choreography, such as the \texttt{VehículoAccidentado} which reports the incident to \texttt{BalizaRole}, and this alerts other vehicles on the road and also publishes the incident to the Central de Balizas, so it can inform the Emergency Center via the "request help" message. This completes the execution as shown in the graph. 

The source code can be downloaded from the following address: \url{https://github.com/GRISE-UPM/ml_server_rest}.

\section{Discussion and conclusions}

The built framework has made it possible to achieve our main work objective: \textit{to apply the existing specifications and standards in SOA for the Coordination of ubiquitous devices in pervasive environments}. The proposed framework works correctly, complying with and respecting the characteristics of ubiquitous devices. The solution found to the research problem is simple, interoperable and extensible, since it is based on an extension of the concepts of a theory that existed only in SOA to an area where they did not exist: the pervasive environments.

The framework presented here uses microservices on the devices, thus constructing an architecture analogous to microservices as a foundation for coordinating them through choreographies. Moreover, both in microservices and in this work, REST mechanisms and XML with JSON as a data exchange format are used. It can also be framed as an improvement to the microservices architecture, as according to Dragoni et al.\cite{dragoni:hal-01631455}, this approach of creating collaborations between microservices through choreographies is of utmost importance for this technology.

This approach has been built on the concepts of SOA, but at the same time, this architecture has some characteristics that make it distinct from SOA, such as: the size of the services, the context in which they execute, the independence of the services, flexibility, modularity, and the evolution that the developed microservices can have. Although it is also important to mention that microservices are not intended to be used in ubiquitous devices, but rather as part of a system based on SOA architecture.

With the advent of microservices technology, \cite{Dragoni2017} points out that the way of composing microservices is primarily based on orchestrations and choreographies. Choreographies are more appropriate, as in these, all components have the same level of relevance and importance in the execution. These concepts are based on the WS-BPEL and WS-CDL standards, respectively, which are derived from SOA.

We are currently working on the refinement and adjustment of a mechanism for the management of transactions within the choreography execution, with a 2 phase commit, through the WS-Transaction specification \cite{OASIS:WSAT:2007}. Another of the current lines  of Investigation is the implementation of a security layer, similar to that of WS-Security \cite{OASIS:WSSECURITY:2006}  and of reliability in both the delivery of messages and the execution as a whole.

There are also other pending works such as the case of a greater parameterization of the framework, to allow new devices to be incorporated into the execution without the need to modify the source code; and support for others communication protocols such as Bluetooth, NFC, etc.

\section{Appends}\label{sec:anexos}

\subsection{Code regarding proof of concept}
\begin{lstlisting}[language=xml , caption={Choreography code in WS-CDL language}, label={fig:coreografia_accidente_wscdl}]
<!-- Choreographies -->
<choreography name="reportarAccidente" root="true">
	<relationship type="tns:Vehiculo_Baliza"/>
	<variableDefinitions>
		<variable name="DatosIncidente" informationType="tns:avisoIncidenteType" mutable="false"/>
		<variable name="Baliza" channelType="tns:BalizasChannel"/>
	</variableDefinitions>
	<sequence>
		<interaction name="reportarAccidente" 
					 channelVariable="tns:Baliza" 
					 operation="informarIncidente" 
					 initiate="true">
			<participate relationshipType="tns:Vehiculo_Baliza" fromRole="tns:VehiculoAccidentadoRole" toRole="tns:BalizaRole" />
			<exchange action="request" name="informarIncidente" informationType="tns:avisoIncidenteType">
				<send variable="cdl:getVariable(tns:DatosIncidente,VehiculoAccidentadoRole)"/>
				<receive variable="cdl:getVariable(tns:DatosIncidente,BalizaRole)"/>
			</exchange>
			<timeout  time-to-complete="PT35S"/>
		</interaction>
		<perform choreographyName="tns:publicarAccidente">
		</perform>
	</sequence>
</choreography>

<choreography name="publicarAccidente" root="false">
	<relationship type="tns:Baliza_CentralBaliza"/>
	<relationship type="tns:Baliza_Vehiculo"/>
	<relationship type="tns:CentralBaliza_CentralEmergencia"/>
	<variableDefinitions>
		<variable name="DatosIncidente" informationType="tns:avisoIncidenteType" mutable="false"/>
		<variable name="CentralBaliza" channelType="tns:CentralBalizasChannel"/>
		<variable name="Vehiculo" channelType="tns:VehiculosChannel"/>
		<variable name="CentralEmergencia" channelType="tns:CentralEmergenciasChannel"/>
	</variableDefinitions>
	<parallel>
		<interaction name="publicarAccidente" 
					 channelVariable="tns:CentralBaliza" 
					 operation="publicarIncidente">
			<participate relationshipType="tns:Baliza_CentralBaliza" fromRole="tns:BalizaRole" toRole="tns:CentralBalizasRole" />
			<exchange action="request" name="informarIncidente" informationType="tns:avisoIncidenteType">
				<send variable="cdl:getVariable(tns:DatosIncidente,BalizaRole)"/>
				<receive variable="cdl:getVariable(tns:DatosIncidente,CentralBalizasRole)"/>
			</exchange>
			<timeout  time-to-complete="PT35S"/>
		</interaction>
		<interaction name="alertaAccidente" 
					 channelVariable="tns:Vehiculo" 
					 operation="alertaIncidente">
			<participate relationshipType="tns:Baliza_Vehiculo" fromRole="tns:BalizaRole" toRole="tns:VehiculoTransitoRole" />
			<exchange action="request" name="informarIncidente" informationType="tns:avisoIncidenteType">
				<send variable="cdl:getVariable(tns:DatosIncidente,BalizaRole)"/>
				<receive variable="cdl:getVariable(tns:DatosIncidente,VehiculoTransitoRole)"/>
			</exchange>
			<timeout  time-to-complete="PT35S"/>
		</interaction>
	</parallel>
</choreography>

<choreography name="solicitarAyudaAccidente" root="false">
	<relationship type="tns:CentralBaliza_CentralEmergencia"/>
	<variableDefinitions>
		<variable name="DatosIncidente" informationType="tns:avisoIncidenteType" mutable="false"/>
		<variable name="CentralEmergencia" channelType="tns:CentralEmergenciasChannel"/>
	</variableDefinitions>
	<workunit name="esperaxcomunicacion"
		guard="cdl:isVariableAvailable(tns:DatosIncidente,CentralBalizasRole)">
			<interaction name="solicitarAyudaAccidente" 
						 channelVariable="tns:CentralEmergencia" 
						 operation="solicitarAyudaIncidente">
				<participate relationshipType="tns:CentralBaliza_CentralEmergencia" fromRole="tns:CentralBalizasRole" toRole="tns:CentralEmergenciasRole" />
				<exchange action="request" name="solicitarAyudaIncidente" informationType="tns:avisoIncidenteType">
					<send variable="cdl:getVariable(tns:DatosIncidente,CentralBalizaRole)"/>
					<receive variable="cdl:getVariable(tns:DatosIncidente,CentralEmergenciasRole)"/>
				</exchange>
				<timeout  time-to-complete="PT35S"/>
			</interaction>
	</workunit>
</choreography>
</package>
\end{lstlisting}

Below is the code for the classes that represent the core of the development, in C++ language, which is the one used in the Arduino boards:

\begin{lstlisting}[language=c++ , caption={Ejemplo de coreografía}, label={fig:ejemplo-coreografia-completo}]
void RESTWebServer::procesarPeticion(const char *input, connection *pServer){
	char verbo;
    bool res = false;
	bool procesado = false;
	int retorno;

    if (input[0] == '\0'){
        return false;
    }

    if (existe_subcadena(input,"GET")){
        res = this->procesar_GET(input);
		verbo = 'G';
    }
    if (existe_subcadena(input,"PUT")){
        res = this->procesar_PUT(input);
		verbo = 'P';
    }
    if (existe_subcadena(input,"POST")){
        res = this->procesar_POST(input);
		verbo = 'O';
    }
    if (existe_subcadena(input,"DELETE")){
        res = this->procesar_DELETE(input);
		verbo = 'D';
    }

    if (res){
		for(int i=0; i <= this->_cntServiciosUtilizados; i++){
			if (strcmp(this->entradas[0],this->serviciosClases[i]) == 0){
				retorno = this->serviciosObj[i]->iniciar(verbo, this->_DATOS, this->cnt_datos,this);
				if (retorno != 1){
					pServer->enviarRespuestaHTTP("503 Service Unavailable","");
					procesado = false;
					break;
				}
				this->serviciosObj[i]->ejecutar(this->entradas[1],this->entradas[2],pServer);
				procesado = true;
			}
		}
	}

    return;
}

bool RESTWebServer::procesar_GET(const char *input){
      // Comienzo el parseo del string de entrada ...
      char *parte, *token;
	  int pos;
      bool res = false;
	  char temp[128];

      cnt_entradas = -1;
      cnt_datos = -1;

	  // Resguardo el string inicial
	  pos = strpos(input,"{");
	  strcpy(temp,input+pos);

      parte = strstr(input, "GET ");
      if (parte != NULL){
		parte = strtok(input,"\n");
		parte = strtok(parte," ");
		parte = strtok(NULL, " ");

        // Primero verifico si esta o no el RewriteBase
        if (existe_subcadena(parte,_RewriteBase)){
            // Parseo todo el string en pedazos de "/"
            token = strtok(parte, "/");
            while(token != NULL){
                entradas[++cnt_entradas] = token;
                if ((cnt_entradas == 0)){
					if (strcmp(strlwr(entradas[cnt_entradas]),_RewriteBase) == 0){
						cnt_entradas--;
					}
                }
                token = strtok(NULL, "/");
            }


            // Intento sacar los parametros del querystring, si es que vinieron
			if (existe_subcadena(parte,"?")){
				parte = strpos(entradas[cnt_entradas], "?") + 1;
				if (strlen(parte) > 0){
					entradas[cnt_entradas] = entradas[cnt_entradas] + strpos(entradas[cnt_entradas],"?");
					token = strtok(parte, "&");
					while(token != NULL){
						_DATOS[++cnt_datos] = token;
						token = strtok(NULL, "&");
					}
				}
			}
            res = true;
        }
    }
    return res;
	}
}
\end{lstlisting}

Here is the code of the \texttt{RESTWebServer} class, responsible for processing the requests \footnote{For space reasons, only the code for processing a \texttt{GET} request is shown.}


\bibliographystyle{ieeetr}
\bibliography{revista.bib}

\end{document}